\documentstyle[prl,preprint,aps,epsf]{revtex}
\begin{document}

\title{Functional Relation of interquark potential with interquark distance}
\author{D.K. Park$^{1,2}$\footnote{e-mail:
dkpark@hep.kyungnam.ac.kr},
H. J. W. M\"{u}ller-Kirsten$^{1}$\footnote{e-mail:
mueller1@physik.uni-kl.de} and 
A. V. Shurgaia $^{1}$\footnote{On leave from A. Razmadze Mathematical
Institute of Georgian Academy of Sciences 380093, Tbilisi Georgia,
email: shurgaia@physik.uni--kl.de}}
\address{1. Department of Physics,
 University of Kaiserslautern, D-67653 Kaiserslautern, Germany\\
2.Department of Physics, Kyungnam University, Masan, 631-701, Korea}

\maketitle

\begin{abstract}
The functional relation between interquark potential
 and interquark distance
is explicitly derived by considering the Nambu--Goto action in the
$AdS_5 \times S^5$ background. It is also shown that a similar 
relation holds in a
general background. The implications of this relation for confinement are 
briefly discussed.
\end{abstract}

\vspace{1cm}

%\centerline{PACS Numbers:04.70.Dy, 04.62.+v, 11.25.-w}

\newpage

%\vspace{3cm}
$AdS/CFT$ correspondence\cite{mal98-1} enables us to understand 
quantum phenomena in the large $N$ limit of ${\cal N}=4$ 
super Yang--Mills theory
with the classical string description\cite{gub98,wit98-1} in the 
$AdS_5 \times S^5$ background. In particular, the finite temperature effect
on the gauge theory side of the correspondence
has been discussed in detail\cite{wit98-2,gross81}, and
on the AdS side
the Wilson loop has been calculated at both
 zero temperature\cite{rey98-1,mal98-2} and
finite temperature\cite{rey98-2,bran98-1}
from worldsheet areas \cite{sonn}. The main differences  
between the finite temperature case
and the
zero temperature case are, in our opinion, 
(1) the presence of a maximal separation distance which makes 
the dependence of the interquark distance on the minimum point of
the string 
a multi--valued function, and
(2) the appearance of a cusp (or 
bifurcation point) in the graph of interquark potential--vs--interquark 
distance\cite{rey98-2}. As in Ref.\cite{mal98-2}
``quark" and ``antiquark" here mean the ends on the
$AdS_5$ boundary of a string describing an infinitely
massive W--boson stretching between
the horizon and the $AdS_5$ boundary. The near--extremal
N D3--branes are located at the Schwarzschild horizon $U=U_T
(r = {\alpha}^{\prime}U_T)$.

The two differences referred to
 strongly suggest that there is a hidden functional
relation between these quantities as we
know from the study of equations of state
in thermal 
physics. 

In the following we derive this relation explicitly in the $AdS_5$
background. Also, it will be shown that a
 similar relation\footnote{For an  arbitrary 
background this is a 
relation between the string energy and the separation of string ends at the
boundary.} holds for an arbitrary background.   

We start with the classical Nambu--Goto action
for the string worldsheet
\begin{equation}
\label{nambu1}
S_{NG} = \frac{1}{2 \pi \alpha^{\prime}}
\int d\tau d\sigma
\sqrt{det G_{MN} \partial_{\alpha} X^M \partial_{\beta} X^N}
\end{equation}
in the near extremal Euclidean Schwarzschild-$AdS_5$ background\cite{mal98-1}
\begin{equation}
\label{ads}
ds_E^2 = \alpha^{\prime}
\left[ \frac{U^2}{R^2} \left( f(U) dt^2 + dx_i dx_i \right) +
\frac{R^2 f(U)^{-1}}{U^2} dU^2 + R^2 d \Omega_5^2 \right].
\end{equation}
Here $x_i, i=1, 2, 3,$ are the  D3--brane coordinates and $R$ is
the radius
of both $AdS_5$ and $S^5$ and $U=r/{{\alpha}^{\prime}}$.
 The function $f(U)$ is defined by
 $f(U) = 1 - {{U_T}^4} / {U^4}$ and
the temperature is given by $T = U_T / (\pi R^2)$ (see e.g.\cite{bran98-1}).
 After identifying 
$\tau = t$ and $\sigma = x$, it is easy to show that for the static case
the action $S_{NG}$ becomes (setting ${\alpha}^{\prime} = 1$)
\begin{equation}
\label{nambu2}
S_{NG} = \frac{\tilde \tau}{2 \pi} \int dx 
\sqrt{U^{\prime 2} + \frac{U^4 - U_T^4}{R^4}} 
\end{equation}
where the prime denotes differentiation with respect to $x$, and 
$\tilde \tau$ is the entire Euclidean time interval.
$U_T$ is the position of the horizon.
Since the Hamiltonian is a constant of motion we have
\begin{equation}
\label{constant}
\frac{U^4 - U_T^4}{\sqrt{U^{\prime 2} + \frac{U^4 - U_T^4}{R^4}}} = 
const \equiv R^2 \sqrt{U_0^4 - U_T^4}
\end{equation}
where the constant solution $U_0$ is the minimum point
 of the string configuration
(by symmetry at $x = 0$). 
One can explicitly
derive the static solution of $S_{NG}$ by integrating eq.(\ref{constant})
 in terms of elliptic functions, i.e.,\cite{byrd1,byrd2} 
\begin{eqnarray}
\label{sol}
x&=& \frac{R^2 \sqrt{U_0^2 - U_T^2}}{2 \sqrt{2} U_0 U_T}
\Bigg[ F\left( \sin^{-1} \frac{\sqrt{(U^2 - U_0^2) (U^2 - U_T^2)}}
                              {U^2 - U_0 U_T},
               \frac{U_0 + U_T}{\sqrt{2 (U_0^2 + U_T^2)}} \right)
                                                            \\   \nonumber 
& &\hspace{3.0cm}
    - F\left( \sin^{-1} \frac{\sqrt{(U^2 - U_0^2) (U^2 - U_T^2)}}
                              {U^2 + U_0 U_T},
               \frac{U_0 - U_T}{\sqrt{2 (U_0^2 + U_T^2)}} \right)
                                                               \Bigg]
\end{eqnarray}
where $F(\phi, k)$ is the incomplete elliptic integral of the first kind
with modulus $k$ of the associated Jacobian elliptic functions.
Since the interquark distance $L$ is defined by the distance
between different ends
of the string on the $AdS_5$ boundary ($U \rightarrow \infty$), it is easy 
to compute $L$ from (\ref{sol}); one finds
(with quark and antiquark at $x=-L/2$ and $L/2$
respectively)
\begin{equation}
\label{distance}
L = \frac{R^2}{\sqrt{2} U_T}
    \frac{\sqrt{a^2 - 1}}{a} \left[ K(f_1(a)) - K(f_2(a)) \right]
\end{equation}
where $a = U_0 / U_T$, $K(k)$ is the complete elliptic integral
of the first kind \cite{byrd1}, and $f_1(a)$ and
$f_2(a)$ are given by
\begin{eqnarray}
\label{f1f2}
f_1(a)&=& \frac{a + 1}{\sqrt{2 (a^2 + 1)}},   \\   \nonumber
f_2(a)&=& \frac{a - 1}{\sqrt{2 (a^2 + 1)}}.
\end{eqnarray}
It is important to note that $f_1(a)$ and $f_2(a)$ are respectively
the modulus and  complementary modulus of the elliptic functions 
 in the sense that $f_1(a)^2 + f_2(a)^2 = 1$. 
Since $U_T$ is the position of the horizon we consider
only $U_0>U_T$, i.e. $a\geq 1$.

The energy of solution (\ref{sol})  is interpreted 
as the interquark potential
in the context of $AdS/CFT$ correspondence, and  
is readily derived using the constant
of motion (\ref{constant}), i.e.,
\begin{equation}
\label{eqqbar}
E_{q\bar{q}} 
=\lim_{{\tilde \Lambda}\rightarrow\infty}\frac{1}{2\pi}\int
^{{\tilde \Lambda}}_{U_0} dx \sqrt{{U^{\prime}}^2+\frac{U^4-U^4_T}{R^4}}
= \frac{U_T}{\pi} \lim_{\Lambda \rightarrow \infty}
\int_a^{\Lambda} d y \sqrt{\frac{y^4 - 1}{y^4 - a^4}}
\end{equation}
where $\Lambda$ is a cutoff parameter. To regularize $E_{q\bar{q}}$,
 we have to
subtract the quark masses\cite{mal98-1}. The finite form of
the interquark 
potential is then
\begin{equation}
\label{finite}
E_{q\bar{q}}^{(Reg)} = \frac{U_T}{\pi} \lim_{\Lambda \rightarrow \infty}
\left[ \int_a^{\Lambda} d y 
\sqrt{\frac{y^4 - 1}{y^4 - a^4}} - (\Lambda - 1) \right].
\end{equation}
This expression can also be evaluated in terms of standard 
elliptic functions\cite{alph}. We obtain
\begin{equation}
\label{analytic}
E_{q\bar{q}}^{(Reg)} = \frac{U_T}{\pi}
\left[ 1 + \sqrt{\frac{a^2 + 1}{2}}
\left[ \frac{a - 1}{2 a} K(f_1(a)) + \frac{a + 1}{2 a} K(f_2(a)) - E(f_1(a))
- E(f_2(a)) \right] \right],
\end{equation}
where $E$ is the complete elliptic integral of the second kind \cite{byrd1}. 

The $L$--dependence of $E_{q\bar{q}}^{(Reg)}$ for various temperatures 
is plotted in 
Fig. 1. As mentioned above, the appearance of cusps at finite temperatures
strongly suggests that there exists a functional relation between 
$E_{q\bar{q}}^{(Reg)}$ and $L$ as we know
from our experience with the thermodynamical treatment of
a van der Waals gas or recent investigation of
systems allowing the tunneling of large spins
\cite{liang1}.
To find this relation it is more convenient to introduce the modulus 
$\kappa = f_2(a) = (a - 1) / \sqrt{2(a^2+1)}$  explicitly.
Since $a \geq 1$\footnote{This is a consequence of Ref.\cite{mal98-3} that
D--branes are located at the horizon. Although we have a somewhat different
opinion, we follow this condition here. 
Our own opinion will be discussed elsewhere.}, $\kappa$ is defined in the 
region $0 \leq \kappa \leq 1 / \sqrt{2}$. Inverting the
relation we have
\begin{equation}
\label{adefine}
a = \frac{1 + 2 \kappa \kappa^{\prime}}{1 - 2 \kappa^2}
\end{equation}
where $\kappa^{\prime} \equiv \sqrt{1 - \kappa^2}$.
Then $L$ and $E_{q\bar{q}}^{(Reg)}$  expressed in terms of 
$\kappa$ and $\kappa^{\prime}$ become:
\begin{eqnarray}
\label{another}
L&=& \frac{R^2}{\sqrt{2} U_T}
\frac{2\sqrt{\kappa \kappa^{\prime} + 2 \kappa^2 \kappa^{\prime 2}}}
     {1 + 2 \kappa \kappa^{\prime}}
    \left[ K(\kappa^{\prime}) - K(\kappa) \right],
                                                   \\   \nonumber
E_{q\bar{q}}^{(Reg)}&=& \frac{U_T}{\pi}
\left[ 1 + \frac{\sqrt{1 + 2 \kappa \kappa^{\prime}}}{1 - 2 \kappa^2}
      \left[ \frac{\kappa (\kappa + \kappa^{\prime})}
                  {1 + 2 \kappa \kappa^{\prime}} K(\kappa^{\prime})
            + \frac{1 + \kappa (\kappa^{\prime} - \kappa)}
                   {1 + 2 \kappa \kappa^{\prime}} K(\kappa)
           - E(\kappa^{\prime}) - E(\kappa) \right] \right].
\end{eqnarray}
Since the thermodynamical relations 
are usually realized on the level of first derivatives,
 we compute $d L / d \kappa$ and $d E_{q\bar{q}}^{(Reg)} / d \kappa$
whose explicit forms are
\begin{eqnarray}
\label{fderivative}
\frac{d L}{d \kappa}&=& \frac{R^2}{\sqrt{2} U_T}
\frac{1}{\kappa^{\prime} (1 + 2 \kappa \kappa^{\prime})
         \sqrt{\kappa \kappa^{\prime} + 2 \kappa^2 \kappa^{\prime 2}}}
                                                      \\   \nonumber
& & \hspace{1.0cm} \times
\left[ (1 + 4 \kappa^3 \kappa^{\prime}) K(\kappa^{\prime}) + (1 + 4 \kappa
     \kappa^{\prime 3}) K(\kappa) - 2 (1 + 2 \kappa \kappa^{\prime})
          \left(E(\kappa) + E(\kappa^{\prime}) \right) \right],
                                                       \\    \nonumber
\frac{d E_{q\bar{q}}^{(Reg)}}{d \kappa}&=& 
\frac{U_T}{\pi} \frac{1}
                     {\kappa^{\prime} (\kappa^{\prime 2} - \kappa^2)^2
                      \sqrt{1 + 2 \kappa \kappa^{\prime}}}
                                                        \\  \nonumber
& & \hspace{1.0cm} \times
\left[ (1 + 4 \kappa^3 \kappa^{\prime}) K(\kappa^{\prime}) + (1 + 4 \kappa
     \kappa^{\prime 3}) K(\kappa) - 2 (1 + 2 \kappa \kappa^{\prime})
          \left(E(\kappa) + E(\kappa^{\prime}) \right) \right].
\end{eqnarray}  
We observe that the coefficients of the complete elliptic 
integrals in the brackets of 
$d L / d \kappa$ and $d E_{q\bar{q}}^{(Reg)} / d \kappa$ are 
identical. This means that $d E_{q\bar{q}}^{(Reg)} / d L$ is independent of 
these complete elliptic integrals, i.e.,
\begin{equation}
\label{eqql1}
\frac{d E_{q\bar{q}}^{(Reg)}}{d L} = 
\frac{\sqrt{U_0^4 - U_T^4}}{2 \pi R^2}.
\end{equation}

The right--hand side of Eq.(\ref{eqql1}) is the regularized energy of 
the constant solution $U = U_0$, i.e.  $E^{(Reg)}(U_0)$.
Thus  we obtain finally
the functional relation
\begin{equation}
\label{final}
\frac{d E_{q\bar{q}}^{(Reg)}}{d L} = E^{(Reg)}(U_0).
\end{equation}
This relation is seen to be very similar to the relation between
dynamical quantities of a classical 
Euclidean point particle in
quantum mechanical tunneling: $d S_E / d P = {\cal E}$ where $S_E$, $P$, 
and ${\cal E}$ are Euclidean action, period and 
energy of the classical particle
(cf. e.g. \cite{liang2}).

It is interesting to compare Eq. (\ref{final}) with the case of a van der
Waals gas\cite{reif}. Our plot of $E_{q\bar{q}}^{(Reg)}$--vs--$L$
 is completely analogous 
to the plot of enthalpy--vs--pressure of the gas whose equation of state
plotted as pressure--vs--volume corresponds to the plot of $L$--vs--$U_0$. 
Hence, the plot of $E_{q\bar{q}}^{(Reg)}$-vs-$L$ is completely determined 
from the plot of $L$-vs-$U_0$ up to the constant. If the plot of 
$L$--vs--$U_0$  has $n$ extremum points, the plot of 
$E_{q\bar{q}}^{(Reg)}$--vs--$L$ has $n$ bifurcation points.

In order to understand our result in greater
 generality we consider the Nambu--Goto action (\ref{nambu1}) in an
arbitrary space--time background. Then the static action will
generally reduce to
\begin{equation}
\label{gennambu}
S_{NG} = \frac{\tilde \tau}{2 \pi} \int dx 
\sqrt{G_1(U) U^{\prime 2} + G_2(U)}
\end{equation}
where $G_1(U)$ and $G_2(U)$ are metric--dependent functions. The constant of 
motion in this case is
\begin{equation}
\label{gconst}
\frac{G_2(U)}{\sqrt{G_1(U) U^{\prime 2} + G_2(U)}} = const
\equiv \sqrt{G_2(U_0)}.
\end{equation}
Hence, we choose $U = U_0$ as an extremum point in the string 
configuration. We assume $G_1(U_0) \neq 0$ and $G_2(U_0) \neq 0$. Although
this assumption excludes particular space-time geometries, we do
not think that 
this assumption restricts our general argument crucially. From the 
constant of motion it is easy to show that the solution of the
static Nambu--Goto action (\ref{gennambu}) obeys the integral equation,i.e.,
\begin{equation}
\label{inteq}
x = \int_{U_0}^{U} d U 
\sqrt{\frac{G_1(U) G_2(U_0)}{G_2(U) [G_2(U) - G_2(U_0)]}}
\end{equation}
>From this integral equation it is straightforward to show that the distance
$L$ between the string ends at $U \rightarrow \infty$ and the string energy
${\cal E}$ become
\begin{eqnarray}
L&=& 2 \sqrt{G_2(U_0)}
    \int_{U_0}^{\infty} d U \sqrt{\frac{G_1(U)}{G_2(U) [G_2(U) - G_2(U_0)]}},
                                                           \\  \nonumber
{\cal E}&=& \frac{1}{\pi} \int_{U_0}^{\infty} d U 
\sqrt{\frac{G_1(U) G_2(U)}{G_2(U) - G_2(U_0)}}.
\end{eqnarray}
Using the Leibnitz chain rule one can prove directly
\begin{eqnarray}
\label{approxi}
\frac{d {\cal E}}{d U_0} &\approx& - 
\frac{\sqrt{G_1(U_0) G_2(U_0)}}{\pi} 
\lim_{U \rightarrow U_0} \frac{1}{\sqrt{G_2(U) - G_2(U_0)}},
                                                    \\  \nonumber
\frac{d L}{d U_0} &\approx& - 2 \sqrt{G_1(U_0)} 
\lim_{U \rightarrow U_0} \frac{1}{\sqrt{G_2(U) - G_2(U_0)}}
\end{eqnarray}
which recovers our relation
\begin{equation}
\label{our}
\frac{d {\cal E}}{d L} = \frac{1}{2 \pi} \sqrt{G_2(U_0)}
\equiv {\cal E}^{(Reg)}(U_0).
\end{equation}

In our opinion, this kind of relation is not restricted to
Nambu--Goto actions. We expect that a similar relation
can be established for a 
Born--Infeld action, which is under study.

Finally we comment briefly on confinement, using our relation (\ref{final}).
Since confinement and deconfinement depend on the $L$--dependence of
$E_{q\bar{q}}^{(Reg)}$, say $E_{q\bar{q}}^{(Reg)} \propto L^{\alpha}$ where 
$\alpha \geq 1$ and $\alpha \leq -1$ represent confinement and 
deconfinement phases respectively, it is important to see the behavior
of 
$d E_{q\bar{q}}^{(Reg)} / d L$ with respect to $L$; this is shown in 
Fig. 2. Each line in Fig. 2 has confining and deconfining 
parts. Fig. 2 also indicates that as the temperature increases, the confining 
force becomes strong (large $\alpha$) while available $L$ is reduced, which
is physically predictable. At low temperature $\alpha$ approaches  $1$ and
hence the area law of the Wilson loop is recovered
in the supergravity picture \cite{bran98-2} (i.e. the $T=0$ 
curve starts at the origin with confining part along
the abscissa, that is, with constant slope
implying $d E_{q\bar{q}}^{(Reg)} / d L$ independent 
of $L$, and as the temperature increases the
confining part has a larger slope, i.e. larger $\alpha$).
It should be noted that the potential energy of the confining part is 
always larger than that of the deconfining part. This
implies that our understanding
of confinement is still incomplete.
A full understanding of quark confinement
 seems to require a deeper 
understanding of the relation between gauge theories and statistical 
physics like that of black holes.

\vspace{3.0cm}
{\bf Acknowledgement:} D.K. P. acknowledges support by the
Deutsche Forschungsgemeinschaft (DFG).

%\newpage
%\centerline{\bf Figure Caption:}
%\vspace{0.4cm}
%\noindent
%{\bf Fig. 1}

%\noindent
\begin{figure}
\caption{
The $L$--dependence of $E_{q\bar{q}}^{(Reg)}$. The appearance of the cusps in 
this figure strongly suggests that there exists a hidden relation between 
$E_{q\bar{q}}^{(Reg)}$ and $L$.}
\end{figure}
\vspace{0.4cm}
%\noindent

%{\bf Fig. 2}

%\noindent
\begin{figure}
\caption{
The $L$--dependence of $d E_{q\bar{q}}^{(Reg)} / d L$. Each line consists of 
a confining (lower) and a deconfining (upper) part. This figure indicates that 
a stronger confining force is required at high temperatures. At  
low temperatures the area law of the Wilson loop is recovered.}
\end{figure}

\newpage
\epsfysize=10cm \epsfbox{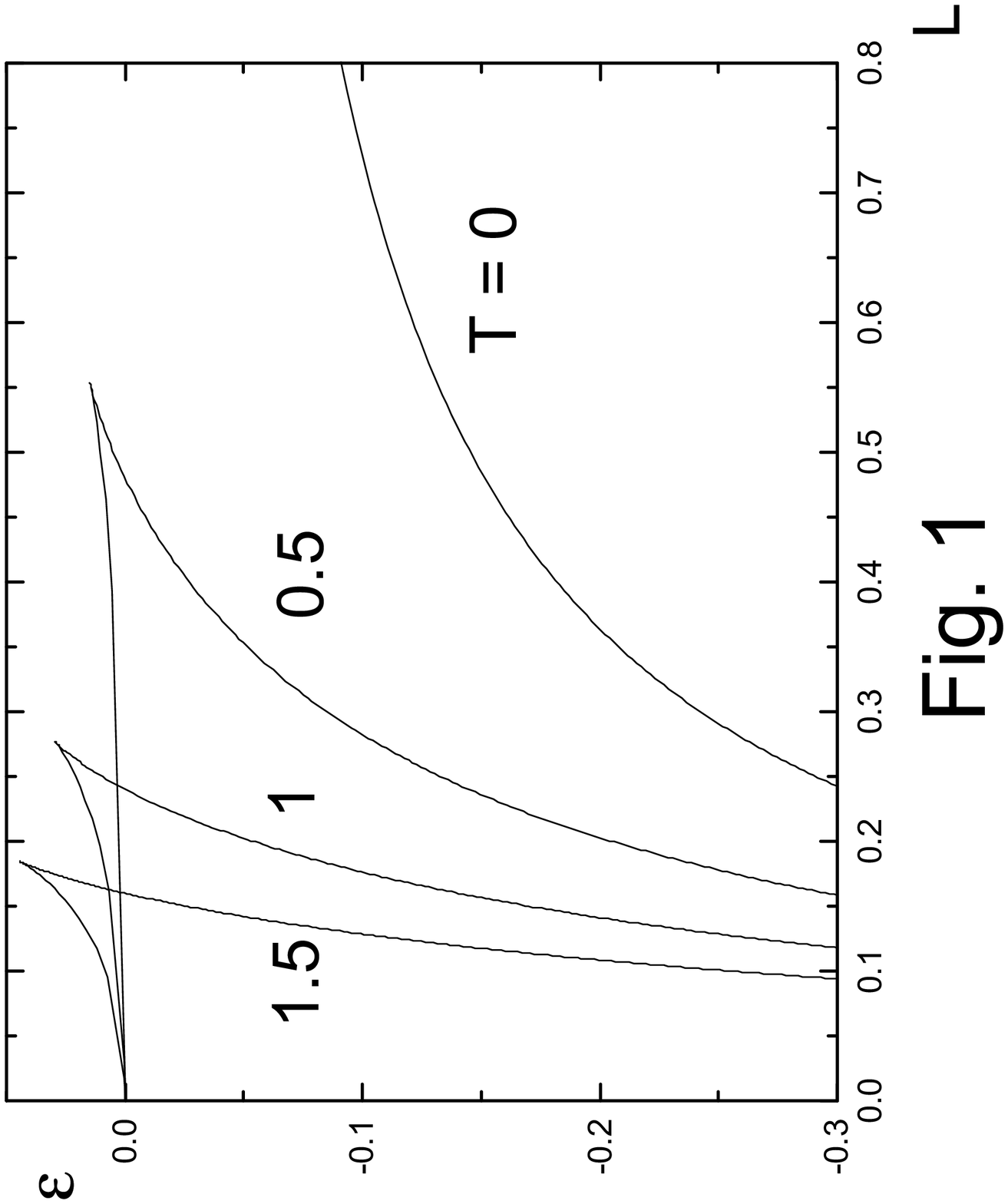}
\newpage
\epsfysize=10cm \epsfbox{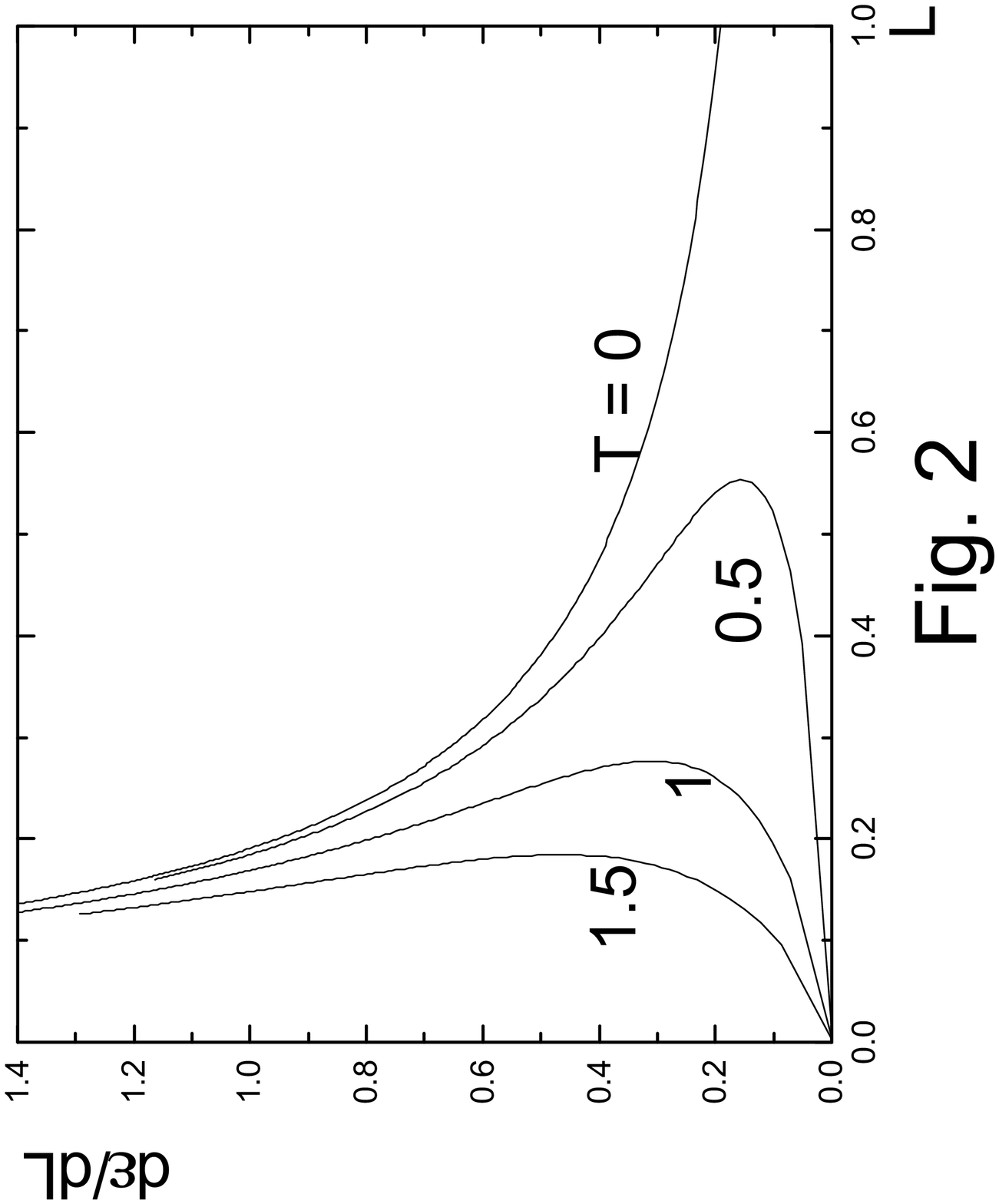}

\end{document}